\documentclass[aps,pra,preprint,groupedaddress]{revtex4-1}

\usepackage{textcomp}
\usepackage{graphicx}
\usepackage{amssymb}
\usepackage{amsmath}

\begin{document}


\title{Young's double-slit interference with two-color biphotons}



\author{De-Jian Zhang$^{1,2}$}
\author{Shuang Wu$^1$}
\author{Hong-Guo Li$^3$}
\author{Hai-Bo Wang$^1$}

\author{Jun Xiong$^{1}$}
\email{junxiong@bnu.edu.cn}

\author{Kaige Wang$^{1}$}
\email{wangkg@bnu.edu.cn}

\address{
$^1$Department of Physics, Applied Optics Beijing Area Major Laboratory,
\\ Beijing Normal University, Beijing 100875, China\\
}
\address{$^2$Department of Physics, Nanchang University, Nanchang 330031, China}
\address{$^3$School of Science, Tianjin University of Technology, Tianjin 300384, China}


\date{\today}

\begin{abstract}
In classical optics, Young's double-slit experiment with colored coherent light gives rise to individual interference fringes for each light frequency, referring to single-photon interference. However, two-photon double-slit interference has been widely studied only for wavelength-degenerate biphoton, known as subwavelength quantum lithography. In this work, we report double-slit interference experiments with two-color biphoton. Different from the degenerate case, the experimental results depend on the measurement methods. From a two-axis coincidence measurement pattern we can extract complete interference information about two colors. The conceptual model provides an intuitional picture of the in-phase and out-of-phase photon correlations and a complete quantum understanding about the which-path information of two colored photons.
\begin{description}
\item[42.50.-p, 42.25.Hz, 42.50.St]
\end{description}
\end{abstract}


\maketitle


\section{Introduction}
Young's double-slit interference is one of the most fundamental and important effects in physics, and the first experimental observation established the wave nature of light. In the early days of quantum mechanics, the phenomenon was regarded as a single-photon effect, and it still is the best example to illustrate wave-particle duality, the uncertainty principle and complementarity. Richard Feynman described the double-slit experiment as one ``which is impossible, absolutely impossible, to explain in any classical way, and which has in it the heart of quantum mechanics. In reality, it contains the only mystery '' \cite{Feynman1965}.

	In the last two decades, an interesting subject appeared regarding two-photon double-slit interference with a quantum entangled light source \cite{Burlakov1997, Hong1998, Fonseca1999a, Fonseca1999b, Kim2000, DAngelo2001, Nogueira2001, Brida2003, Shimizu2003, Kim2011, Walborn2010}  and a thermal light source \cite{Wang2004, Scarcelli2004, Xiong2005, Zhai2005}. In this case, at least two photons pass through the double-slit at the same time. The interference fringes can be observed in the two-photon coincidence counts, i.e. through second-order correlation measurements, at two detectors behind the double-slit, while interference does not appear in the single-photon measurement at either of the two detectors. In the experiments using a quantum entangled source, a two-photon entangled state (biphoton) generated by spontaneous parametric down-conversion (SPDC) is employed. When the coincidence measurement is performed by fixing one detector and scanning the other, the interference fringe spacing is the same as that of single-photon interference \cite{Burlakov1997}. However, when the two detectors are moved together in the same direction, the fringe spacing of the two-photon interference pattern is one half of that for the single-photon case \cite{Fonseca1999a, DAngelo2001, Shimizu2003}. This effect is regarded as subwavelength interference in the sense that a two-photon de Broglie wave  has been created with half the original wavelength \cite{Boto2000, Edamatsu2002}. Similar phenomena are observed for a thermal light source when the two detectors are moved in opposite directions during the intensity correlation measurement \cite{Scarcelli2004, Xiong2005, Zhai2005}.

	Two-photon interference of a quantum entangled source can be interpreted with the concept of biphoton coherence. A pair of entangled photons produced in the SPDC process forms a biphoton, which plays the role as an independent entity with apparent ``first-order coherence'', although each individual photon does not satisfy the first-order coherence. As for a thermal light source, the interference information can be extracted through the intensity correlation measurements. A theoretical comparison between the two sources was performed in Ref. \cite{Wang2004}.

	In this article we report on a Young's double slit interference experiment with two photons of different color. The passage from degenerate photon pair to two-color pair represents a substantial step forward, since the photons can now be identified by their color and the previous quantum interpretation is not appropriate. It is necessary to face some crucial problems. When the two photons passing through the double slit have different colors, one may even doubt that interference is possible at all. If it is possible, how the two different frequencies work to generate the interference pattern? And how is the matter about the which-path information when the two photons can be identified by their color? These questions have not been addressed in previous experimental studies using two-photon with two colors with a beam splitter or a Mach-Zehnder interferometer \cite{Ou1988, Ou1988a, Rarity1990, Larchuk1993, Shih1994, Kim2003, Kim2003a, Liu2012}.

\section{Theoretical Analysis}
\begin{figure}[htb]
\includegraphics[width=8.4cm]{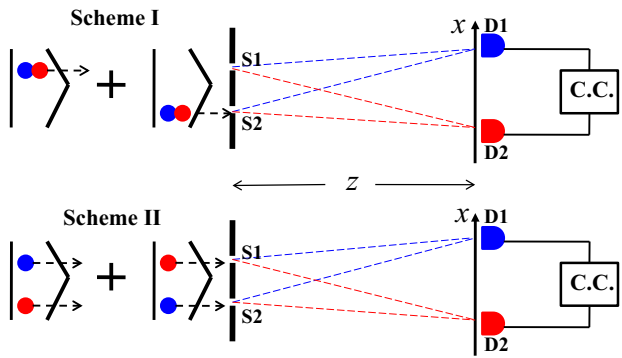}
\caption{Two schemes of two-photon double-slit interference. C.C. denotes the coincidence counting.\label{fig:scheme} }
\end{figure}

The entangled photon pairs produced in SPDC consist of signal and idler photons with different wavelengths, satisfying momentum conservation. Unlike single-photon interference, there are two configurations in which two-photon double-slit interference can occur according to the way that the two photon pairs travel through the double-slit. As shown schematically in Fig.\ref{fig:scheme}, in Scheme I the two photons in each pair pass together through either one of the two slits, while in Scheme II they pass separately through the slits. Previous studies on double-slit interference with frequency degenerate photon pairs focused mainly on Scheme I \cite{Burlakov1997, Fonseca1999a, Fonseca1999b, Kim2000, DAngelo2001, Nogueira2001, Shimizu2003, Kim2011, Walborn2010}, while only some considered Scheme II \cite{Hong1998, Brida2003}. In the present experiment, however, the two photons are identified by their color at the two detectors D1 and D2.

	The two schemes involve two kinds of biphoton entangled states at the double-slit
\begin{subequations}
\begin{equation}
 \left|\Psi_{\mathrm{I}}\right\rangle = \frac{1}{\sqrt{2}}[a_1^\dag(\omega_1)a_1^\dag(\omega_2)+a_2^\dag(\omega_1)a_2^\dag(\omega_2)]\left|0\right\rangle,
\end{equation}
\begin{equation}
 \left|\Psi_{\mathrm{II}}\right\rangle = \frac{1}{\sqrt{2}}[a_1^\dag(\omega_1)a_2^\dag(\omega_2)+a_1^\dag(\omega_2)a_2^\dag(\omega_1)]\left|0\right\rangle,
\end{equation}
\end{subequations}
where $a_1^\dag$ and $a_2^\dag$ are the field creation operators at slits S1 and S2, respectively. At the observation plane, the two detectors D1 and D2 are covered by filters F1 and F2 to select the frequencies $\omega_1$ and $\omega_2$, respectively. Although each detector can identify the photon color, it does not know which slit the photon came from. As a result, the fields recorded by the detectors D1 and D2 are, respectively,
\begin{subequations}
\begin{equation}
  E^{(+)}_1(x_1) = a_1(\omega_1)\exp(i\omega_1r_{11}/c) + a_2(\omega_1)\exp(i\omega_1r_{21}/c) \label{eq:2a},
\end{equation}
\begin{equation}
  E^{(+)}_2(x_2) = a_1(\omega_2)\exp(i\omega_2r_{12}/c) + a_2(\omega_2)\exp(i\omega_2r_{22}/c) \label{eq:2b},
\end{equation}
\end{subequations}
where $r_{ij}$ ($i,j=1,2$) is the distance between slit S$i$ and detector D$j$, $x_i$ is the position of detector D$i$ at the observation plane, and $c$ is the speed of light.

    The coincidence counting rates at the two detectors for the two schemes are calculated to be
\begin{subequations}
\begin{eqnarray}
  & &\left\langle\Psi_{\mathrm{I}}\right| E^{(-)}_1(x_1)E^{(-)}_2(x_2)E^{(+)}_2(x_2)E^{(+)}_1(x_1) \left|\Psi_{\mathrm{I}}\right\rangle \\ \nonumber
  &\propto& 1 + \cos[(\omega_1/c)(r_{21}-r_{11})+(\omega_2/c)(r_{22}-r_{12})] \\ \nonumber
  &=& 1 + \cos[(\omega_1x_1+\omega_2x_2)d/(cz)] \\ \nonumber
  &=& 1 + \cos\{[\omega_+(x_1+x_2)/2+\omega_-(x_1-x_2)/2]d/(cz)\},\label{eq:3a}
\end{eqnarray}
\begin{eqnarray}
  & &\left\langle\Psi_{\mathrm{II}}\right| E^{(-)}_1(x_1)E^{(-)}_2(x_2)E^{(+)}_2(x_2)E^{(+)}_1(x_1) \left|\Psi_{\mathrm{II}}\right\rangle \\ \nonumber
   &\propto& 1 + \cos[(\omega_1/c)(r_{21}-r_{11})-(\omega_2/c)(r_{22}-r_{12})] \\ \nonumber
  &=& 1 + \cos[(\omega_1x_1-\omega_2x_2)d/(cz)] \\ \nonumber
  &=& 1 + \cos\{[\omega_+(x_1-x_2)/2+\omega_-(x_1+x_2)/2]d/(cz)\},\label{eq:3b}
\end{eqnarray}
\end{subequations}
where $d$ is the slit spacing, and $z$ the distance between the double-slit and observation plane. $\omega_+=\omega_1+\omega_2$ is the sum-frequency, and $\omega_-=\omega_1-\omega_2$ is the difference-frequency.

\begin{figure}[htb]
\includegraphics[width=15cm]{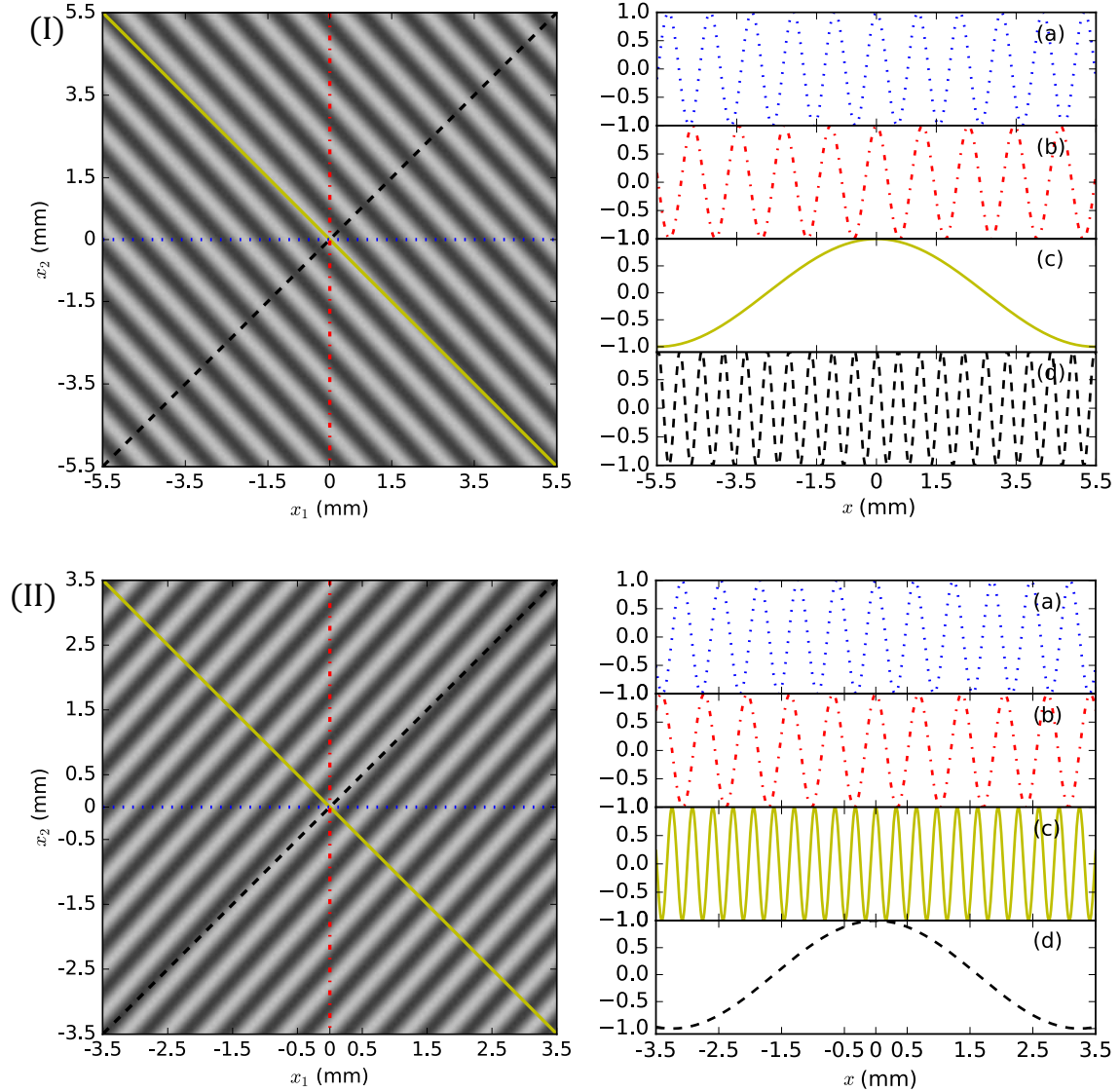}
\caption{Numerically plotted intensity correlation patterns as a function of positions $x_1$ and $x_2$ (left part) for two-color biphotons and their corresponding interference fringes (right part). Figures (I) and (II) represent Schemes I and II, respectively, and figures (a)-(d) are the intensity correlation fringes of the four different detection ways, i.e. for: (a) $\omega_1$ when $x_2=0$ and $x_1=x$; (b) $\omega_2$ when $x_1=0$ and $x_2=x$; (c) $\omega_-$ for Scheme I and $\omega_+$ for Scheme II when $x_1=-x_2=x$; (d) $\omega_+$ for Scheme I and $\omega_-$ for Scheme II when $x_1=x_2=x$.\label{fig:theoretical_results}}
\end{figure}

	All the physical effects of biphoton double-slit interference can be revealed in a two-axis ($x_1$ by $x_2$) intensity correlation (coincidence count) interference pattern, which is plotted theoretically in Fig.\ref{fig:theoretical_results}. The parameters used in this theoretical simulation come from the experimental setups described in the section below. For the frequency degenerate case, the fringe stripes for both Schemes I and II are exactly along the anti-diagonal and diagonal directions, respectively. However, they will deviate from these directions for the frequency non-degenerate case, as shown in Fig.\ref{fig:theoretical_results}. There are four possible measurement outcomes, designated by four colored lines in the contour patterns, which manifest the interference fringes at four frequencies: $\omega_1$ and $\omega_2$ for the two nondegenerate photons, the sum-frequency $\omega_+$ and the difference-frequency $\omega_-$. When detector D2 (D1) is fixed and D1 (D2) is scanned, the interference pattern for frequency $\omega_1$($\omega_2$) is represented by the blue dotted (red dash-dotted) lines for both schemes. However, when both detectors are moved together in the same direction, as shown by the diagonal (black dashed lines) in Fig.\ref{fig:theoretical_results}, we should see the interference fringe of the sum-frequency $\omega_+$ for Scheme I and difference-frequency $\omega_-$ for Scheme II. When the detectors are moved in opposite directions, as shown by the anti-diagonal (yellow solid lines) in Fig. \ref{fig:theoretical_results}, we see the interference fringe of the difference-frequency $\omega_-$ for Scheme I and sum-frequency $\omega_+$ for Scheme II. These results can be readily understood from Eq. (3).

\section{Experiment}
\begin{figure}[htb]
\includegraphics[width=8.4cm]{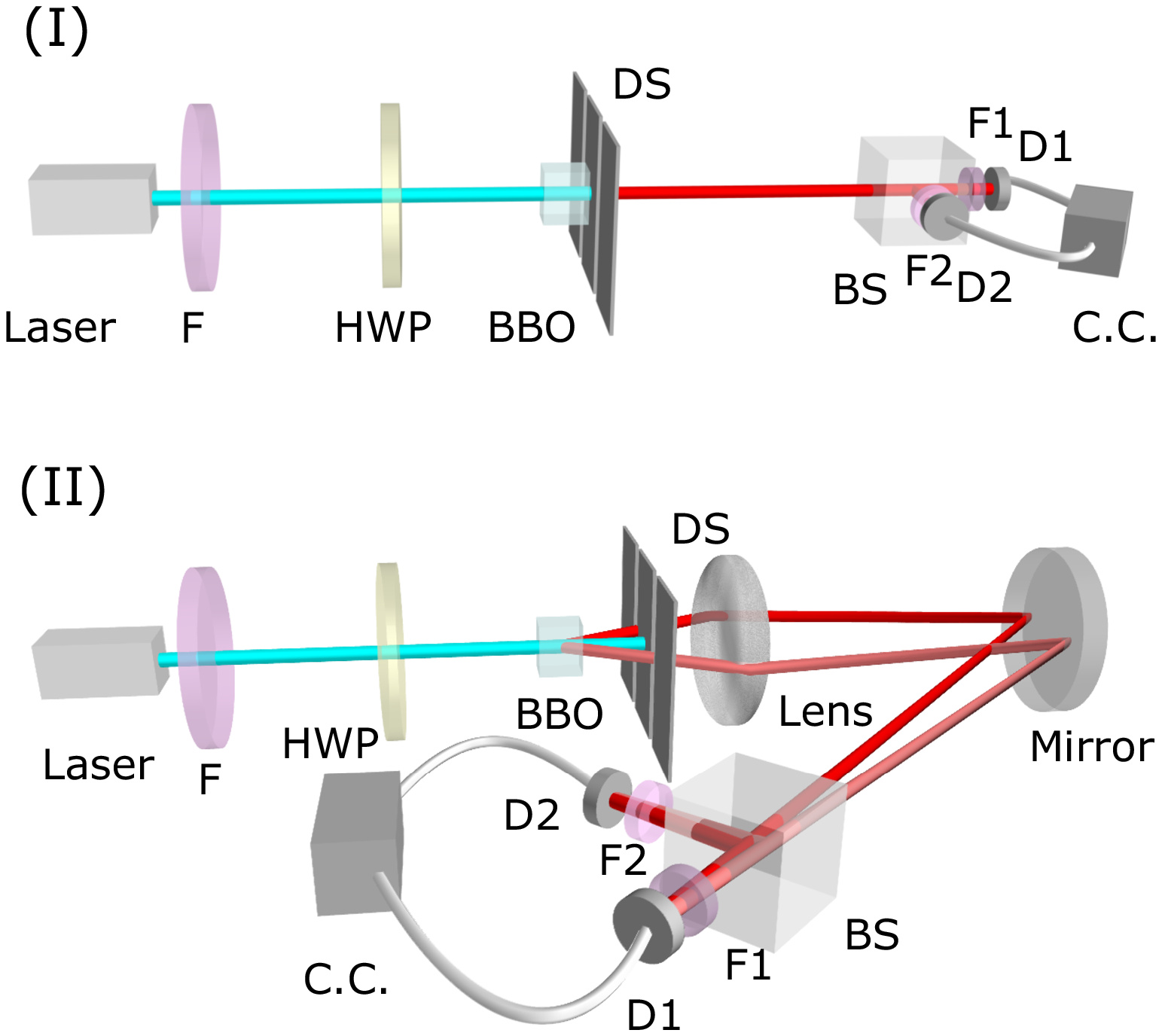}
\caption{Experimental setups of biphoton double-slit interference for Schemes I and II. HWP is the half-wave plate, DS the double-slit, BS the non-polarization 50/50 beam splitter. D1 and D2 are two single-photon detectors. F, F1 and F2 are filters.\label{fig:setup}}
\end{figure}

The experimental setups are shown in Figs.\ref{fig:setup}(I) and \ref{fig:setup}(II) for Schemes I and II, respectively. The entangled photon pairs are obtained by pumping a 5mm$\times$5mm$\times$3mm type I phase matched beta-barium-borate (BBO) crystal with a 330 mW pulsed UV laser beam of 400 nm wavelength and 76 MHz repetition rate, obtained by frequency-doubling a Ti:sapphire femtosecond laser (Mira-900 Coherent, Inc.). After passing through the double-slit, the photon pairs are split into two beams with a non-polarizing 50-50 beam splitter. The transmitted and reflected beams then pass through narrow-band-filters of 10nm full-width-half-max to be detected by single-photon detectors D1 and D2 which are mounted on translation stages.

	For Scheme I, the double-slit is placed right after the BBO crystal. Since the entangled photon pairs are generated at the same location of the crystal, each pair can go through the same slit at any time. The detection plane is 54.7 cm away from the double-slit. As for Scheme II, the BBO crystal is slightly tilted to guarantee that the down-converted signal beam and idler beam diverge at a certain angle. The double-slit is placed at a distance of 1.1 cm from the crystal, so that signal and idler photons pass separately through the two slits. To obtain a high coincidence counting rate at the detection plane, a lens of focal length 6 cm is placed 5.1 cm away behind the double-slit, and the detection plane is at a distance 188 cm from the lens. The system has an effective diffraction length of 30.3 cm. The same double-slit of slit spacing $d=400 \mu$m and slit width $b=100 \mu$m is employed in both schemes.

\begin{figure}[htb]
\includegraphics[width=8.4cm]{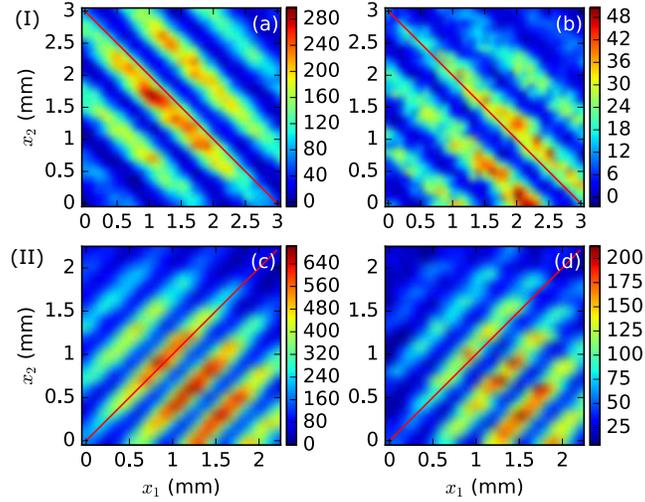}
\caption{Experimental results of the two-axis coincidence count interference patterns: (a) and (b) for Scheme I, (c) and (d) for Scheme II. (a) and (c) show the 800nm frequency degenerate case; (b) and (d), 760nm-840nm non-degenerate case. Color bars show coincidence counts.\label{fig:results_2d}}
\end{figure}

	The experimental results of the two-axis coincidence count (intensity correlation) interference patterns are plotted in Fig.\ref{fig:results_2d}, where Figs. \ref{fig:results_2d}(a)-\ref{fig:results_2d}(b) show the patterns for Scheme I, and Figs.\ref{fig:results_2d}(c)-\ref{fig:results_2d}(d), for Scheme II. For comparison, Figs.\ref{fig:results_2d}(a) and \ref{fig:results_2d}(c) show the interference patterns for the frequency degenerate case, where filters of 800 nm are placed in front of both detectors. When the 760 nm and 840 nm filters are used, the two-axis patterns for the non-degenerate case are shown in Figs.\ref{fig:results_2d}(b) and \ref{fig:results_2d}(d). The experimental outcomes demonstrate that the interference fringes for the degenerate photon pairs lie exactly along the diagonals, while the fringes for the non-degenerate photon pairs are slightly (about $3$ degrees) off from the diagonal directions.

\begin{figure}[htb]
\includegraphics[width=15cm]{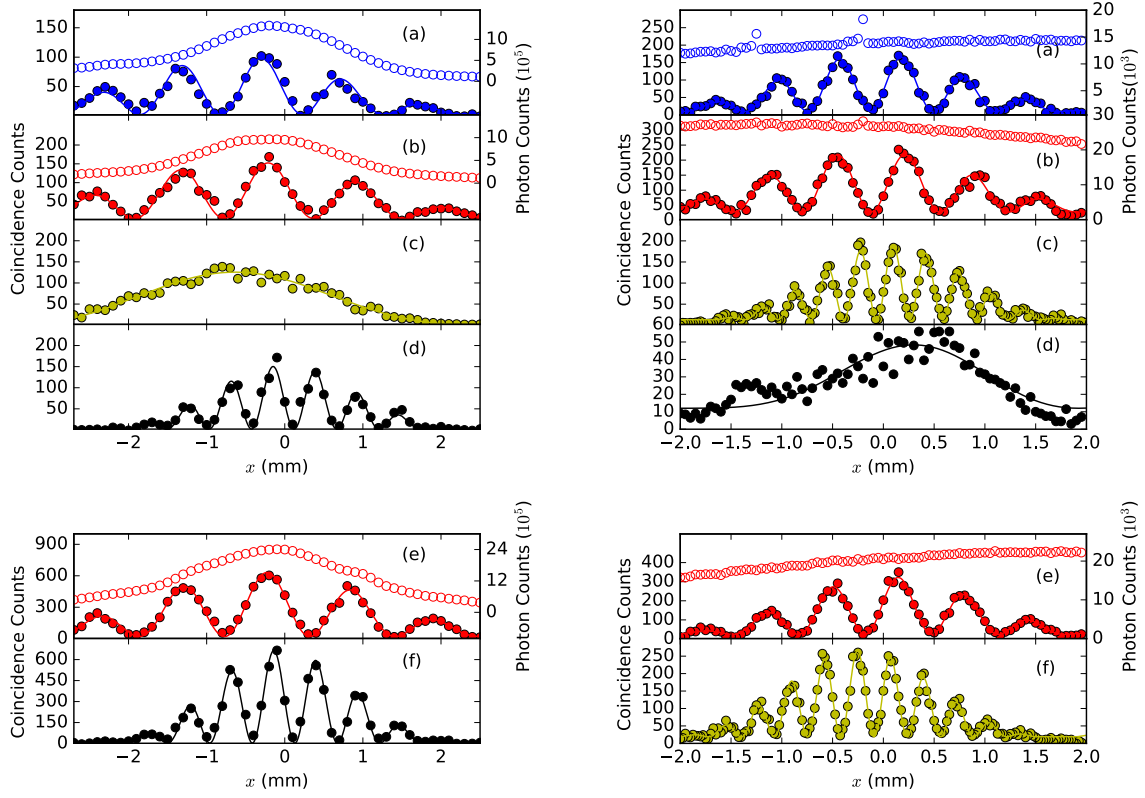}
\caption{Experimental results of interference fringes for Scheme I (left) and Scheme II (right): (a)-(d) are for two colored biphotons of wavelengths 760 and 840nm; (e) and (f) are for degenerate biphotons of 800nm. Four detection outcomes, shown by different colors, are the same as in Fig. 2. Open-circles are single-photon counts, and full-circles are coincidence counts.\label{fig:results_1d}}
\end{figure}

	The one-dimensional interference patterns are shown in Fig.\ref{fig:results_1d} where the left part is for Scheme I and the right part for Scheme II. For the two colored photons of frequencies 760nm and 840nm, the interference fringes reconstructed from the four observation ways are shown in Figs.\ref{fig:results_1d}(a)-\ref{fig:results_1d}(d). When only one detector is scanned in the coincidence measurement, the interference fringes of individual single-photon frequencies are plotted in Fig.\ref{fig:results_1d}(a) and Fig.\ref{fig:results_1d}(b) for both schemes. A slight difference in the fringe spacing for the two colors can be seen. However, when both detectors are moved in the same or opposite directions, sum-frequency interference fringes are observed in the left Fig.\ref{fig:results_1d}(d) and right Fig.\ref{fig:results_1d}(c) figures, respectively. On the other hand, in the opposite observation configuration, the difference-frequency fringes are not observed in either the left Fig.\ref{fig:results_1d}(c) or right Fig.\ref{fig:results_1d}(d) figures, since in our present experiments the interference range at the detection plane is less than the fringe spacing of the difference-frequency interference. As a matter of fact, evidence of non-degenerate two-photon interference has already appeared in Fig.\ref{fig:results_2d}. For comparison, the fringes for the degenerate two-photon case are recorded in Fig.\ref{fig:results_1d}(e) and Fig.\ref{fig:results_1d}(f) for both the schemes. In all the cases above, the single-photon counting results do not show any interference effect. The experimental data are in good agreement with the theoretical simulation.

\begin{figure}[htb]
\includegraphics[width=8.4cm]{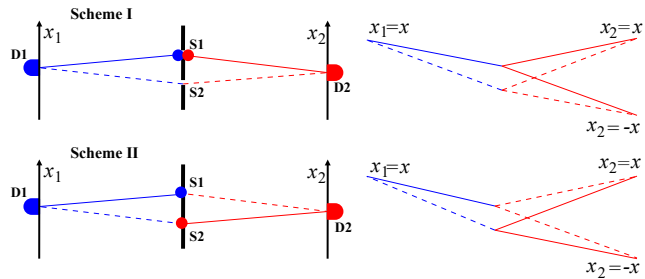}
\caption{Sketch of the conceptual model for Scheme I (upper) and Scheme II (lower).\label{fig:equimodel}}
\end{figure}

\section{Discussions}
The double-slit interference with two-color biphotons can be understood by a model which is similar to those proposed in Refs.\cite{DAngelo2001, Wang2004}. In the conceptual model of Fig.\ref{fig:equimodel}, one of the detectors is reset to a position symmetric to the double-slit, while either of the two detectors can be regarded as a source emitting a photon at the corresponding frequency. The two-photon interference occurs between two paths: one denoted by two solid lines and the other by two dashed lines between D1 and D2. For Scheme I, source D1 emits a blue photon and travels to slit S1 (or S2), where the blue photon converts immediately to a red photon and continuously travels to detector D2. As for Scheme II, source D1 emits a blue photon and travels to slit S1 (or S2), then a red photon, immediately created at the other slit S2 (or S1) to replace the blue one, travels to detector D2. When D1 as the blue source is fixed, detector D2 records the interference pattern for the red photon, equivalent to one-photon double-slit interference. The same case occurs when D2 is fixed and D1 records one-photon double-slit interference for the blue beam.

	To observe the sum- and difference-frequency interference patterns, the source and the detector in the conceptual model must move synchronously in same or opposite directions. In the right part of Fig.\ref{fig:equimodel}, we can see that the phase difference of the two paths between D1 and D2 is the superposition of the two phase differences for blue and red photons. We define in-phase as both dashed lines having longer (or shorter) paths than the corresponding solid lines, while in the out-of-phase case, one dashed line is longer and the other is shorter than the corresponding solid lines. In Scheme I, when $x_1=x_2=x$ ($x_1=-x_2=x$), the two phase differences for blue and red photons are in-phase (out-of-phase), resulting in two-photon sum-frequency (difference-frequency) interference. The same illustration applies to Scheme II.

	Just as for the single-photon double-slit interference, in Scheme I we do not have any which-path information for the biphotons. In Scheme II, however, we do know that each photon passes through a single-slit each time but we still have no knowledge of its color. When the two photons are degenerate and unentangled in Scheme II, the two-photon interference becomes the Hanbury-Brown and Twiss type interference of two independent photons \cite{Wang2004}. In this case, the detectors are still unable to tell which slit the photon came from. As a matter of fact, in all the cases, the detectors can identify the photon color but can never know which path it took.

\section{Conclusions}
	In conclusion, the double-slit interference with two-color biphoton exhibits more interesting phenomena than other biphoton interference experiments, and may afford a complete understanding of two-photon interference. We proposed a two-axis interference pattern to replace the conventional interference fringe in Young's double-slit experiment, and this is necessary to describe complete interference information about two-color frequencies. We can now understand that the subwavelength interference observed in previous studies is actually the result of sum-frequency interference for degenerate photon pair while the difference-frequency effect disappears. The conceptual model has manifested that the interference of two-color biphoton can be regarded as the correlation of two colored photons, embodying the nonlocality nature of quantum entanglement. The work provides an intuitive picture of in-phase and out-of-phase non-degenerate photon correlations, and a complete and general description about the which-path information of two colored photons in quantum interference.

\section{Acknowledgments}
We thank L. A. Wu for helpful discussions. This work was supported by the National Natural Science Foundation of China, Projects No. 11474027, No. 61675028 No. 11174038, and the National High Technology Research and Development Program of China, Project No. 2013AA122902.

D. J. Z. and  S. W.  contributed equally to this work.




%

\end{document}